\documentclass[aps,prl,twocolumn,showpacs]{revtex4}

\usepackage{epsfig}
\usepackage{graphicx}
\usepackage{amsmath}
\usepackage{bm}

\begin{document}

\title{Strength of the $d_{x^2-y^2}$ pairing in the two-leg Hubbard ladder}

\author{N. Bulut and S. Maekawa}

\address{Institute for Materials Research, Tohoku University, Sendai
980-8577, Japan \\
CREST, Japan Science and Technology Agency (JST), Kawaguchi,
Saitama 332-0012, Japan}

\date{August 21, 2005}

\begin{abstract}
In the ground state of the doped two-leg Hubbard ladder there are
power-law decaying $d_{x^2-y^2}$-type pairing correlations. It is
important to know the strength and the temperature scale of these
correlations. For this purpose, we have performed determinantal
Quantum Monte Carlo (QMC) calculations of the reducible
particle-particle interaction in the Hubbard ladder. In this
paper, we report on these calculations and show that, at
sufficiently low temperatures, resonant particle-particle
scattering takes place in the $d_{x^2-y^2}$ pairing channel for
certain values of the model parameters. The QMC data presented
here indicate that the $d_{x^2-y^2}$ pairing correlations are
strong in the Hubbard ladder.
\end{abstract}


\maketitle

It is generally accepted that the high-$T_c$ cuprates
\cite{Bednorz} are $d_{x^2-y^2}$-wave superconductors
\cite{Scalapino,Tsuei}. It is also known that the spin-fluctuation
exchange in electronic models \cite{MSV,SLH} and an
electron-phonon interaction which has its largest coupling at
small wavevectors \cite{Zeyher,Bulut96,Huang} can lead to an
effective electron-electron interaction which is attractive in the
$d$-wave Bardeen-Cooper-Schrieffer (BCS) pairing channel. However,
it is not yet clear what is the pairing interaction responsible
for superconductivity in the high-$T_c$ cuprates. Within this
context, it is important to determine the maximum possible
strength of the $d_{x^2-y^2}$ pairing correlations which one can
obtain in a purely electronic model such as the Hubbard model.
Unfortunately, the ground state of the two-dimensional (2D)
Hubbard model remains beyond the reach of the exact many-body
techniques \cite{Bulut}. However, in the case of the two-leg
Hubbard ladder, the Density Matrix Renormalization Group (DMRG)
calculations found power-law decaying $d_{x^2-y^2}$-type
pair-field correlations \cite{Noack94,Noack97}. This is probably
the only model where it is known from exact calculations that the
pairing correlations get enhanced by turning on an onsite Coulomb
repulsion in the ground state \cite{Yamaji94,Noack97}. Hence, it
is important to determine the maximum possible strength and the
temperature scale of the $d_{x^2-y^2}$ pairing correlations in the
Hubbard ladder, which is the main motivation of this paper.

Here, we present Quantum Monte Carlo (QMC) results on the
reducible particle-particle interaction $\Gamma$ in the BCS
channel, which is illustrated in Fig. 1. The reducible $\Gamma$
serves as a powerful probe of the pairing correlations. For
example, in the case of an $s$-wave superconductor, $\Gamma$ at
the Fermi surface would diverge to $-\infty$ due to repeated
particle-particle scatterings in the BCS channel, when the
superconducting transition is approached, $T\rightarrow T_c^+$. In
this paper, we investigate the strength of the $d_{x^2-y^2}$
pairing in the Hubbard ladder by making use of the exact QMC data
on $\Gamma$.

In the following, we show that the reducible particle-particle
interaction $\Gamma$ in the Hubbard ladder exhibits diverging
behavior as the temperature is lowered for certain values of the
model parameters. We find that, on the Fermi surface, $\Gamma$ can
become strongly repulsive (attractive) for ${\bf q}\approx
(\pi,\pi)$ (${\bf q}\approx 0$) momentum transfers, which
correspond to backward (forward) scatterings. In particular, near
half-filling and for Coulomb repulsion $U=4t$ and temperature
$T=0.1t$, where $t$ is the hopping matrix element, the backward
and forward scattering amplitudes can become an order of magnitude
larger than the bare Coulomb repulsion or the bare bandwidth. This
type of momentum dependence of $\Gamma$ implies that resonant
particle-particle scattering is taking place in the BCS channel
already at $T\approx 0.1t$. For $U=8t$, we observe similar
behavior at about twice higher temperatures. The temperatures
studied in this paper are lower than those reached in previous QMC
calculations of $\Gamma$ for the 2D \cite{Bulut93} and the two-leg
\cite{Dahm97} Hubbard models. In addition, we present results on
the solution of the Bethe-Salpeter equation in the BCS channel,
which quantitatively determines the strength of the pairing
correlations. The QMC data shown in this paper imply that the
$d_{x^2-y^2}$-type pairing in the two-leg Hubbard ladder is strong
for certain values of the model parameters.

\begin{figure}
\centering
\includegraphics[height=2.0cm,
bbllx=0,bblly=0,bburx=455,bbury=126,clip]{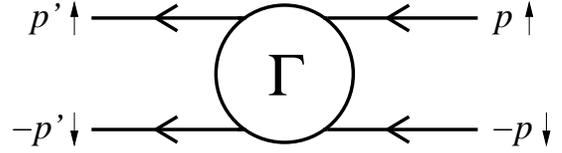} \caption{ Feynman
diagram for the reducible particle-particle interaction
$\Gamma(p'|p)$ in the BCS channel. Here, $p$ denotes $({\bf
p},i\omega_n)$ with Matsubara frequency $\omega_n$. In this
diagram, the incoming fermions at states $p$ with up spin and $-p$
with down spin scatter to states $p'$ with up spin and $- p'$ with
down spin by exchanging $q=p'-p$.} \label{fig1}
\end{figure}
We begin by briefly describing the previous studies of
$d_{x^2-y^2}$ pairing in the Hubbard ladder. The DMRG calculations
found that the rung-rung pair-field correlation function decays as
power law in the ground state of the doped Hubbard ladder
\cite{Noack94}. The mean-field calculations suggested the
$d_{x^2-y^2}$ type of symmetry for pairing in doped spin ladders
\cite{Gopalan}. The exact-diagonalization \cite{Yamaji94} and the
DMRG \cite{Noack97} calculations found that the $d_{x^2-y^2}$
pairing correlations are most enhanced when the interchain hopping
$t_{\perp}$ is greater than the intrachain hopping $t$, in
particular, for $t_{\perp}\approx 1.5t$ in the intermediate
coupling regime and near half-filling. In this case, the QMC
calculations showed that the irreducible particle-particle
interaction peaks for momentum transfers near $(\pi,\pi)$ due to
antiferromagnetic fluctuations \cite{Dahm97}.

The two-leg Hubbard model is defined by
\begin{eqnarray}
\label{Hubbard} H= &&-t \sum_{i,\lambda,\sigma}
(c^{\dagger}_{i,\lambda,\sigma}c_{i+1,\lambda,\sigma} + {\rm
h.c.}) - t_{\perp}\sum_{i,\sigma}
(c^{\dagger}_{i,1,\sigma}c_{i,2,\sigma} \nonumber \\  &&+{\rm
h.c.}) + U \sum_{i,\lambda} n_{i,\lambda,\uparrow}
n_{i,\lambda,\downarrow} -\mu \sum_{i,\lambda,\sigma}
n_{i,\lambda,\sigma},
\end{eqnarray}
where $t$ is the hopping parameter parallel to the chains (along
${\bf \hat{x}}$), and $t_{\perp}$ is for hopping perpendicular to
the chains (along ${\bf \hat{y}}$). The operator
$c^{\dagger}_{i,\lambda,\sigma}$ ($c_{i,\lambda,\sigma}$) creates
(annihilates) an electron of spin $\sigma$ at site $i$ of chain
$\lambda$, and
$n_{i,\lambda,\sigma}=c_{i,\lambda,\sigma}^{\dagger}c_{i,\lambda,\sigma}$
is the electron occupation number. As usual, $U$ is the onsite
Coulomb repulsion, and $\mu$ is the chemical potential. In
addition, periodic boundary conditions were used along the chains,
and $t_{\perp}$ denotes interchain hopping for open boundary
conditions.

In obtaining the data presented here, the determinantal QMC
technique described in Ref.~\cite{White89} was used. The
calculation of the reducible interaction $\Gamma$ follows the
procedure described in Ref.~\cite{Bulut93} for the 2D case. The
BCS component of the reducible particle-particle interaction
$\Gamma(p'|p)$ is illustrated in Fig. 1, where $p=({\bf
p},i\omega_n)$ with Matsubara frequency $\omega_n=(2n+1)\pi T$. In
the following, results will be shown for the reducible interaction
in the singlet channel, $\Gamma_{\rm s} (p'|p) = {1 \over 2} \big[
\Gamma(p'|p) + \Gamma(-p'|p) \big]$. In particular, the momentum
dependence of $\Gamma_{\rm s}({\bf p'},i\omega_{n'}|{\bf
p},i\omega_n)$ will be shown at the lowest Matsubara frequency
$\omega_n=\omega_{n'}=\pi T$. In the ground state and for $U=4t$,
the $d_{x^2-y^2}$-type pairing correlations are most enhanced near
half-filling for $t_{\perp}\approx 1.6t$ \cite{Noack97}. When
$U=8t$, this occurs for $t_{\perp}\approx 1.4t$. In this paper,
the QMC data will be presented using these two parameter sets for
a $2\times 16$ lattice.

\begin{figure}[t]
\centering
\includegraphics[height=8.1cm,
bbllx=28,bblly=155,bburx=569,bbury=662,clip]{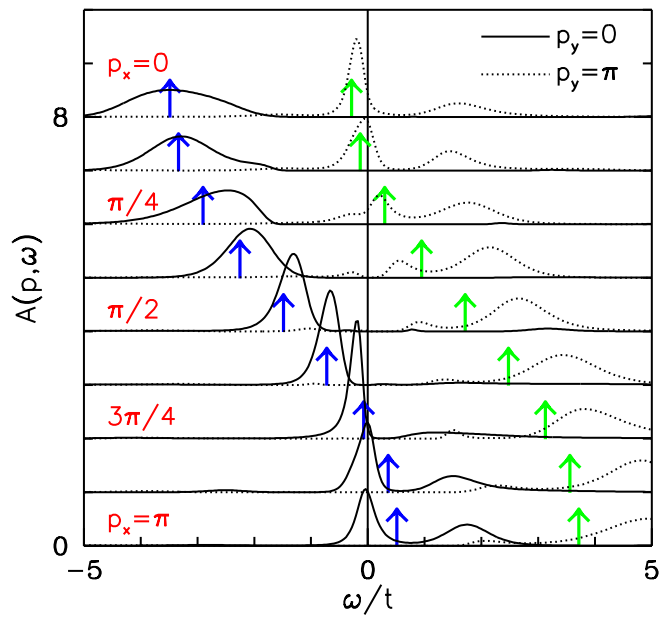} \caption{
(color online) Single-particle spectral weight $A({\bf p},\omega)$
versus $\omega$ at various ${\bf p}$.
The solid and dotted curves
represent the bonding ($p_y=0$) and antibonding ($p_y=\pi$)
bands, respectively.
These results are for $T=0.1t$, $U=4t$,
$t_{\perp}=1.6t$ and $\langle n\rangle=0.94$.
The arrows denote the quasiparticle positions for the $U=0$ case.
} \label{fig2}
\end{figure}
The momentum structure in $\Gamma_{\rm s}(p'|p)$ depends
sensitively on where ${\bf p'}$ and ${\bf p}$ are located with
respect to the Fermi surface. Hence, we will first show results on
the single-particle spectral weight $A({\bf p},\omega)=-
\frac{1}{\pi} {\rm Im}\,G({\bf p},\omega+i\delta)$, where $G$ is
the single-particle Green's function, and discuss the location of
the Fermi-surface crossing points. We have obtained $A({\bf
p},\omega)$ from the QMC data on the single-particle Green's
function along the Matsubara-time axis by using the
maximum-entropy analytic continuation method. Figure 2 shows
$A({\bf p},\omega)$ versus $\omega$ for $T=0.1t$, $U=4t$,
$t_{\perp}=1.6t$ and $\langle n\rangle=0.94$. Here, we see that
the Fermi level crossing for the antibonding ($p_y=\pi$) band
occurs for $p_x$ between $\pi/8$ and $\pi/4$, while for the
bonding ($p_y=0$) band there is spectral weight pinned near the
Fermi level for $3\pi/4 \leq p_x \leq \pi$ at this temperature.
These results for $A({\bf p},\omega)$ are similar to those
presented in Refs. \cite{Noack97,Dahm97}. In the following,
$\Gamma_{\rm s}({\bf p'},i\pi T | {\bf p},i\pi T)$ versus ${\bf
p}$ will be shown for ${\bf p'}=(\pi/4,\pi)$ near the Fermi level
and for ${\bf p'}=(0,\pi)$ at the saddle point.

\begin{figure}[t]
\centering
\includegraphics[height=8.6cm,
bbllx=115,bblly=150,bburx=453,bbury=718,angle=-90,clip]{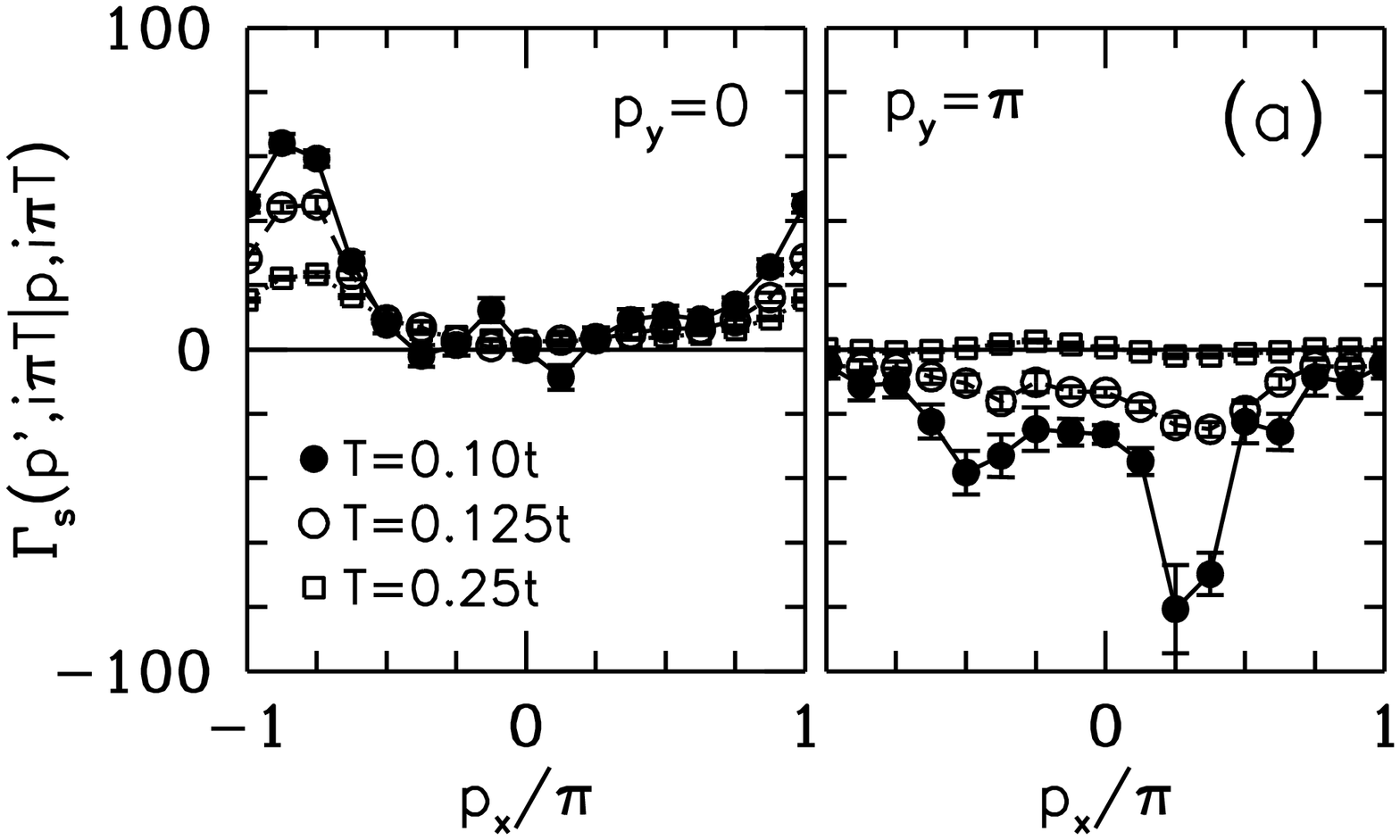}\\
\includegraphics[height=8.6cm,
bbllx=115,bblly=150,bburx=453,bbury=718,angle=-90,clip]{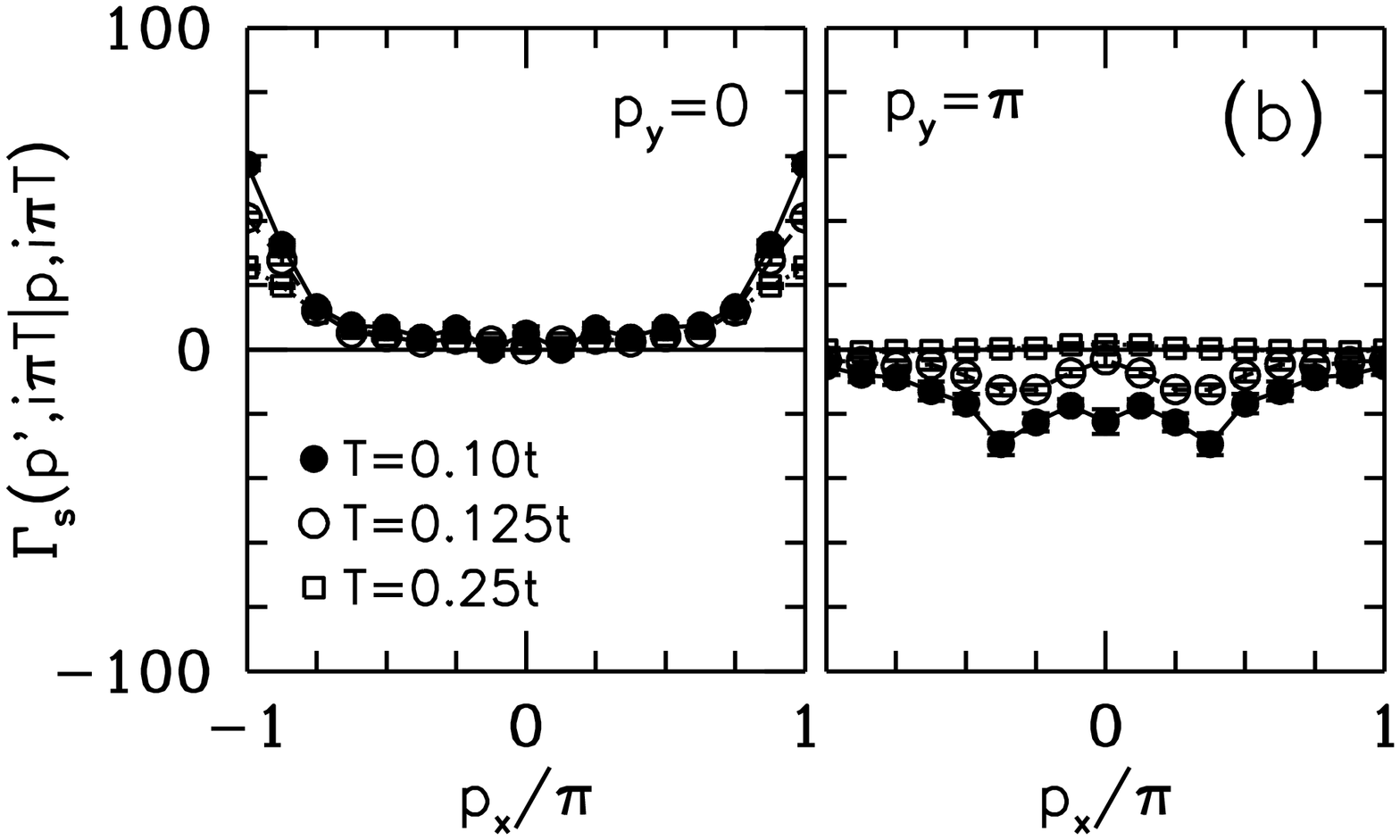}
\caption{Reducible particle-particle interaction in the singlet
channel $\Gamma_{\rm s}({\bf p'},i\pi T|{\bf p},i\pi T)$ versus
$p_x$.
Here, $p_x$ is scanned for $p_y=0$ (left panel) and for $p_y=\pi$
(right panel).
In (a), ${\bf p'}$ is kept fixed at $(\pi/4,\pi)$,
while in (b), ${\bf p'}=(0,\pi)$.
Here, $\Gamma_{\rm s}$ is shown in units of $t$
for $U=4t$, $t_{\perp}=1.6t$ and $\langle n\rangle=0.94$.
} \label{fig3}
\end{figure}
Figure 3(a) shows $\Gamma_{\rm s}({\bf p'},i\pi T | {\bf p},i\pi
T)$ versus ${\bf p}$ while ${\bf p'}$ is kept fixed at
$(\pi/4,\pi)$. In the left panel, $p_x$ is scanned from $-\pi$ to
$\pi$ for $p_y=0$, while in the right panel $p_x$ is scanned for
$p_y=\pi$. Here, it is seen that repulsive and attractive peaks
develop in $\Gamma_{\rm s}$, as $T$ decreases from $0.25t$ to
$0.1t$. In particular, in the left panel it is seen that, when
${\bf p}\approx (-3\pi/4,0)$, a repulsive peak develops in
$\Gamma_{\rm s}$, which corresponds to a scattering process with
momentum transfer ${\bf q}\approx(\pi,\pi)$. In addition, in the
right panel it is observed that a dip develops in $\Gamma_{\rm s}$
when ${\bf p}\approx {\bf p'}=(\pi/4,\pi)$ corresponding to zero
momentum transfer. This dip is due to resonant scattering in the
$d_{x^2-y^2}$-wave BCS channel. In a three-dimensional infinite
system, when a $d_{x^2-y^2}$-wave superconducting instability is
approached, $\Gamma_{\rm s}({\bf p'},i\pi T|{\bf p},i\pi T)$ at
the Fermi level diverges to $+\infty$ for backward scattering, and
to $-\infty$ for forward scattering, which will be further
discussed below. In Fig. 3(a), it is seen that $\Gamma_{\rm s}$ is
developing this type of repulsive and attractive peaks at $T$ of
order $0.1t$.

Figure 3(b) shows $\Gamma_{\rm s}({\bf p'},i\pi T| {\bf p},i\pi
T)$ versus ${\bf p}$ while ${\bf p'}$ is kept fixed at the saddle
point $(0,\pi)$. In this case, $\Gamma_{\rm s}$ develops a peak
when ${\bf p}= (\pm\pi,0)$, corresponding to scattering with ${\bf
q}=(\pi,\pi)$ momentum transfer. The magnitude of this peak is
comparable to that seen in the left panel of Fig.~3(a). However,
the behavior for ${\bf q}=0$ momentum transfer is different. As
observed in the right panel of Fig.~3(b), $\Gamma_{\rm s}$ for
${\bf q}=0$ scattering remains pinned near zero for $T$ down to
$0.125t$, and becomes attractive only below this temperature.
Hence, at the saddle point $(0,\pi)$,
the resonant scattering in $\Gamma_{\rm s}$ for ${\bf q}=0$
momentum transfer is weaker compared to that at $(\pi/4,\pi)$.

\begin{figure}
\centering
\includegraphics[height=4.5cm,
bbllx=61,bblly=173,bburx=512,bbury=539,clip]{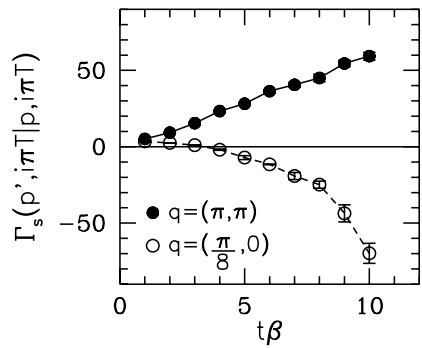} \caption{
Reducible particle-particle interaction in the singlet channel
$\Gamma_{\rm s}({\bf p'},i\pi T|{\bf p},i\pi T)$ versus the
inverse temperature $\beta$ for backward (${\bf q}=(\pi,\pi)$) and
forward (${\bf q}=(\pi/8,0)$) scatterings. Here, ${\bf p}={\bf
p'}+{\bf q}$ and ${\bf p'}$ is kept fixed at $(\pi/4,\pi)$. These
results are for $U=4t$, $t_{\perp}=1.6t$ and $\langle
n\rangle=0.94$. } \label{fig4}
\end{figure}
Figure 4 shows the $T$ dependence of the backward and
forward scattering components of $\Gamma_{\rm s}$. Here,
$\Gamma_{\rm s}$ is plotted as a function of the inverse
temperature $\beta$ for momentum transfers ${\bf q}=(\pi,\pi)$
and $(\pi/8,0)$, while ${\bf p'}$ is kept fixed at $(\pi/4,\pi)$.
This figure shows that, near the Fermi level,
the backward and forward scattering
amplitudes increase rapidly at low $T$, becoming an order of
magnitude larger than the bare Coulomb repulsion at
$T\approx 0.1t$.

Figures 3 and 4 display the main features of $\Gamma_{\rm s}$,
which can be summarized as follows: At low $T$, there are strong
${\bf q}\approx(\pi,\pi)$ scatterings over the whole Brillouin
zone, while the ${\bf q}\approx 0$ scatterings are most attractive
near the Fermi surface. These features were observed at $\langle
n\rangle=0.94$ and 0.875, and for $U=4t$ and $8t$. The one-loop
Renormalization-Group (RG) technique is also utilized to obtain
the momentum dependence of $\Gamma$ in the Hubbard model
\cite{Furukawa,Honerkamp}. It would be useful to make comparisons
of the QMC and the RG results for the momentum dependence of
$\Gamma$ in the ladder case.

In order to determine the strength of the pairing correlations,
the Bethe-Salpeter equation for the reducible particle-particle
interaction in the BCS channel,
\begin{equation}
\label{BS} {\lambda_{\alpha} \over 1-\lambda_{\alpha}}
\phi_{\alpha}(p) = - {T \over N} \sum_{p'} \, \Gamma(p|p')
|G(p')|^2 \phi_{\alpha}(p'),
\end{equation}
was solved for $\lambda_{\alpha}$ and the corresponding
eigenfunctions $\phi_{\alpha}({\bf p},i\omega_n)$ by using QMC
data on $\Gamma$ and the single-particle Green's function $G(p)$.
For a three-dimensional infinite system, when the maximum
$\lambda_{\alpha}$ reaches 1, this signals a BCS instability to a
state where the pair wave function has the form of the
corresponding eigenfunction. For a one-dimensional system,
$\lambda_{\alpha}$'s will always be less than 1, however, the $T$
dependence of $\lambda_{\alpha}$'s gives information about the
characteristic temperature scale of the pairing correlations. For
instance, if the maximum eigenvalue reaches 0.9 at some
temperature, then this means that the leading pairing correlations
are enhanced by a factor of 10 through repeated particle-particle
scatterings in the BCS channel. Hence, at this temperature, the
system would exhibit strong pairing fluctuations.
\begin{figure}
\centering
\includegraphics[height=8.6cm,
bbllx=131,bblly=156,bburx=452,bbury=714,angle=-90,clip]{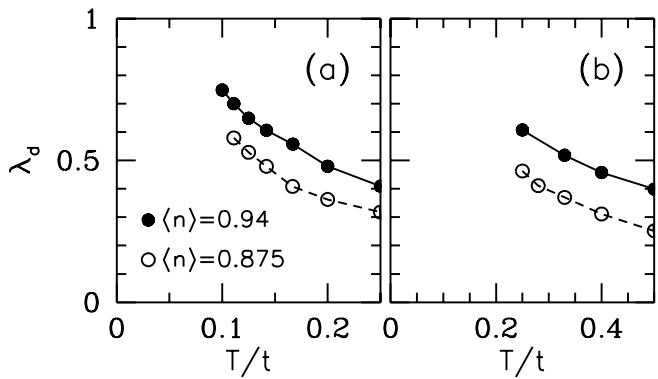}
\caption{$d$-wave irreducible eigenvalue $\lambda_d$ of the
Bethe-Salpeter equation versus $T$ (a) for $U=4t$ and
$t_{\perp}=1.6t$, and (b) for $U=8t$ and $t_{\perp}=1.4t$. Here,
the errorbars are smaller than the size of the symbols. }
\label{fig5}
\end{figure}

At low temperatures, the maximum $\lambda_{\alpha}$ of the
Bethe-Salpeter equation corresponds to an eigenfunction
$\phi_d({\bf p},i\omega_n)$ which has $d_{x^2-y^2}$ type of
symmetry in the sense that it changes sign as ${\bf p}$ goes from
$(\pi,0)$ to $(0,\pi)$ in the Brillouin zone \cite{Dahm97}. The
temperature evolution of the $d_{x^2-y^2}$-wave irreducible
eigenvalue $\lambda_d$ is shown in Figure 5(a) for $U=4t$ and
$t_{\perp}=1.6t$. As $T$ decreases, $\lambda_d$ grows
monotonically reaching 0.75 at $T=0.1t$ for $\langle
n\rangle=0.94$. The simple extrapolation of these results suggests
that $\lambda_d$ will reach 0.9 at $T>0.05t$ for $\langle
n\rangle=0.94$. At the temperatures where these calculations were
performed, $\lambda_d$ decreases upon doping to $\langle
n\rangle=0.875$. Because of the QMC ``fermion sign problem"
\cite{Loh}, these data were obtained by using parallel computers.
Also shown in Fig. 5(b) is $\lambda_d$ for $U=8t$ and
$t_{\perp}=1.4t$. We find that, at $T\geq 0.25t$, $\lambda_d$
takes larger values for $U=8t$ than for $U=4t$. These results
indicate that the $d_{x^2-y^2}$ pairing correlations are strong in
the Hubbard ladder.

An important feature of the QMC data presented here is that, at
sufficiently low $T$, $\Gamma_{\rm s}$ near the Fermi level
becomes strongly attractive for ${\bf q}\approx 0$ momentum
transfers. This is due to resonant scattering in the
$d_{x^2-y^2}$-wave BCS channel. In order to demonstrate this
effect, consider the case of an irreducible interaction
$\Gamma_{\rm I}$ which is independent of frequency and separable
in momentum,
\begin{equation}
\Gamma_{\rm I}({\bf p'}|{\bf p}) = \sum_{\alpha} V_{\alpha}
g_{\alpha}({\bf p'}) g_{\alpha}({\bf p})
\end{equation}
where $\alpha$ denotes the various pairing channels.
In this case, the reducible interaction is given by
\begin{equation}
\label{V}
\Gamma({\bf p'}|{\bf p}) = \sum_{\alpha} {V_{\alpha} \over 1 -
V_{\alpha} P_{\alpha}} g_{\alpha}({\bf p'})
g_{\alpha}({\bf p}),
\end{equation}
with $P_{\alpha} = -(T/N) \sum_p \, g_{\alpha}^2({\bf p})
|G(p)|^2$. In general, $\Gamma_{\rm I}$ has both attractive and
repulsive $V_{\alpha}$. In Eq.~(\ref{V}), it is seen that the
repulsive components get suppressed by repeated scatterings in the
BCS channel, while the attractive components get enhanced. For the
$d_{x^2-y^2}$ channel, $g_d({\bf p})={1\over 2} (\cos p_x - \cos
p_y)$. If the $d_{x^2-y^2}$ component of $\Gamma_{\rm I}$ is
attractive and gets sufficiently enhanced, here it is observed
that $\Gamma$ for ${\bf q}={\bf p'}-{\bf p}\approx 0$ can become
attractive. When a $d_{x^2-y^2}$-wave BCS instability is
approached, then, at the Fermi surface, $\Gamma_{\rm s}$ diverges
to $-\infty$ for ${\bf q}\approx 0$ scatterings and to $+\infty$
for ${\bf q}\approx(\pi,\pi)$ scatterings. The important point is
that the exact QMC data for $\Gamma_{\rm s}$ also exhibit this
type of momentum dependence in the Hubbard ladder at temperatures
which are not low.

In summary, we have presented QMC data on the reducible
particle-particle interaction $\Gamma_s$ in order to determine the
strength and the temperature scale of the $d_{x^2-y^2}$ pairing
correlations in the two-leg Hubbard ladder. We found that
$\Gamma_s$ displays diverging behavior as the temperature
decreases. In particular, $\Gamma_s$ exhibits resonant scattering
in the $d_{x^2-y^2}$ pairing channel as it would be expected for a
system near a $d_{x^2-y^2}$ BCS instability. We have also shown
results on the solution of the Bethe-Salpeter equation to
quantitatively determine the strength of the $d_{x^2-y^2}$
pairing. The QMC data presented in this paper indicate that the
$d_{x^2-y^2}$ pairing correlations in the Hubbard ladder are
strong for certain values of the model parameters. We note that,
for comparison, it would be useful to investigate the strength of
$d_{x^2-y^2}$ pairing for a two-leg ladder system which includes
both onsite Coulomb and electron-phonon interactions.

The authors thank C. Honerkamp, Z.X. Shen, and T. Tohyama
for helpful discussions.
Part of the numerical calculations reported
in this paper were carried out at the ULAKBIM High Performance
Computing Center at the Turkish Scientific and Technical Research
Council. One of us (N.B.) would like to thank the
International Frontier Center for Advanced Materials at Tohoku
University for its kind hospitality, and gratefully acknowledges
support from the Japan Society for the Promotion of Science
and the Turkish Academy of Sciences (EA-TUBA-GEBIP/2001-1-1).
This work was supported by Priority-Areas Grants from the Ministry of
Education, Science, Culture and Sport of Japan, NAREGI Japan and
NEDO.


\end{document}